\documentclass[a4paper,11pt]{article}
\usepackage{pos}

\title{ PeV Gamma-ray Astronomy With Panoramic Optical SETI Telescopes}
 \ShortTitle{PeV Gamma-ray Astronomy With PANOSETI Telescopes}

\author*[a]{Nikolas Korzoun}
\author[b]{Wystan Benbow}
\author[c]{Aaron Brown}
\author[a]{Gregory Foote}
\author[b]{William F. Hanlon}
\author[d]{Olivier Hervet}
\author[d,e]{John Hoang}
\author[a]{Jamie Holder}
\author[f]{Paul Horowitz}
\author[e,g]{Wei Liu}
\author[c]{Jérôme Maire}
\author[e,g]{Nicolas Rault-Wang}
\author[e,g]{Dan Werthimer}
\author[c,h]{James Wiley}
\author[d]{David A. Williams}
\author[c,h]{Shelley A. Wright}

\affiliation[a]{Department of Physics and Astronomy and the Bartol Research Institute, University of Delaware,\\
  104 The Green, Newark, DE 19716, USA}
\affiliation[b]{Center for Astrophysics $|$ Harvard \& Smithsonian,\\
  60 Garden St, Cambridge, MA 02138, USA}
\affiliation[c]{Department of Astronomy \& Astrophysics, University of California San Diego,\\
  9500 Gilman Dr, La Jolla, CA 92093, USA}
\affiliation[d]{Santa Cruz Institute for Particle Physics and Department of Physics, University of California,\\
  552 Red Hill Rd, Santa Cruz, CA 95604, USA}
\affiliation[e]{Department of Astronomy, University of California Berkeley,\\
  501 Campbell Hall, Berkeley, CA 94720, USA}
\affiliation[f]{Department of Physics, Harvard University,\\
  17 Oxford St, Cambridge, MA 02138, USA}
\affiliation[g]{Space Sciences Laboratory, University of California Berkeley,\\
  7 Gauss Way, Berkeley, CA 94720, USA}
\affiliation[h]{Department of Physics, University of California San Diego,\\
  9500 Gilman Dr, La Jolla, CA 92093, USA}

\emailAdd{nkorzoun@udel.edu}

\abstract{The Panoramic Search for Extraterrestrial Intelligence (PANOSETI) experiment is designed to detect pulsed optical signals on nanosecond timescales. PANOSETI is therefore sensitive to Cherenkov radiation generated by extensive air showers, and can be used for gamma-ray astronomy. Each PANOSETI telescope uses a 0.5 m Fresnel lens to focus light onto a 1024 pixel silicon photomultiplier camera that images a 9.9$^\circ$$\times$9.9$^\circ$ square field of view. Recent detections of PeV gamma-rays from extended sources in the Galactic Plane motivate constructing an array with effective area and angular resolution surpassing current observatories. The PANOSETI telescopes are much smaller and far more affordable than traditional imaging atmospheric Cherenkov telescopes (IACT), making them ideal instruments to construct such an array. We present the results of coincident observations between two PANOSETI telescopes and the gamma-ray observatory VERITAS, along with simulations characterizing the performance of a PANOSETI IACT array.}

\ConferenceLogo{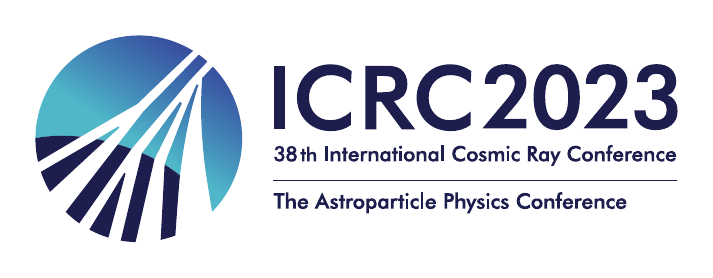}

\FullConference{%
38th International Cosmic Ray Conference (ICRC2023)\\
  26 July - 3 August, 2023\\
  Nagoya, Japan}

\begin{document}
\maketitle

\section{Introduction}
In 2021, LHAASO announced the detection of ultrahigh-energy (UHE) photons ($>$ 0.1 PeV) from 12 sources along the Galactic Plane \cite{cao_ultrahigh-energy_2021}. Of these UHE sources, only the Crab Nebula was firmly established as a counterpart to one of the emitters. With the release of their first gamma-ray catalog, LHAASO now report UHE emission from 43 sources \cite{2023arXiv230517030C}. Observatories like LHAASO and HAWC  are uniquely suited to survey a majority of the sky for UHE emission due to their wide fields of view, comprehensive duty cycles, and large effective areas \cite{2023arXiv230517030C,Albert_2020}. These properties together provide sensitivity to the low flux of UHE photons (5 photons km$^{-2}$ day$^{-1}$ at 0.1 PeV) \cite{cao_ultrahigh-energy_2021}. Many of the new LHAASO sources are spatially extended, and 32 are unassociated with other objects (5 of which are UHE) \cite{2023arXiv230517030C}. Resolving these sources will require targeted studies with instruments that can provide superior angular resolution to LHAASO and HAWC. Arrays of IACTs benefit from increased angular resolution at high energies as a result of stereoscopic reconstruction of the imaged air showers, making them a suitable alternative to particle detectors. The challenge is to design an experiment that makes use of the IACT technique and also meets the effective area requirements in a way that is cost effective and easy to deploy. 

We propose an excellent way to achieve this goal is to use the technology developed by PANOSETI, an experiment designed to search for both optical SETI targets and optical astrophysical transients in underexplored, short time-domains. The plan is to simultaneously observe the whole sky on nanosecond timescales, using many small and easily affordable telescopes \cite{2018SPIE10702E..5IW}. This makes PANOSETI  sensitive to Cherenkov photons produced by atmospheric air showers. The telescopes house four quadrant boards, each with an array of four 8$\times$8 silicon photomultiplier (SiPM) square pixels for a total camera size of 32$\times$32 pixels. Each SiPM array (Hamamatsu S13361-3050AE-08) can resolve single photons on the order of a nanosecond. Custom readout boards shape the signal for triggering an image with precise timing. Light is focused on the camera through a 0.5 meter, acrylic Fresnel lens (Orafol SC214) \cite{2018SPIE10702E..5LM}. The total field of view is 9.9$^\circ$ on a side, giving a pixel width of roughly 0.31$^\circ$ \cite{2018SPIE10702E..5HC}. By positioning a cluster of telescopes together and pointing them in different directions, a "fly's eye" dome can view the whole sky at once \cite{2018SPIE10702E..5LM}. Two such domes separated by about a kilometer would be ideal for observing pulsed, point-like signals. The same telescope components can instead be used on steerable mounts separated by a few hundred meters to observe Chernekov showers. Telescopes are currently installed at two locations at Lick Observatory in California \cite{2018SPIE10702E..5HC}, with two more sites being considered for a pathfinder array that would study other astrophysical phenomena. Here we consider UHE gamma-ray astronomy as one such application. 

To emphasize how PANOSETI's design contrasts with contemporary IACTs, consider the specifications of VERITAS, an array of four IACTs located at the Fred Lawrence Whipple Observatory (FLWO) in southern Arizona. Each telescope is 12 m in diameter, with a camera composed of 499 photomultiplier tubes that together image a total field of view of 3.5$^\circ$. The VERITAS energy sensitivity ranges from 100 GeV to >30 TeV \cite{2015ICRC...34..771P}. To further appreciate how two disparate telescopes can facilitate the same science, we test the performance of PANOSETI telescopes in conjunction with VERITAS.

\section{Coincident Observing}
\begin{figure}
    \centering
    \includegraphics[width=0.67\textwidth]{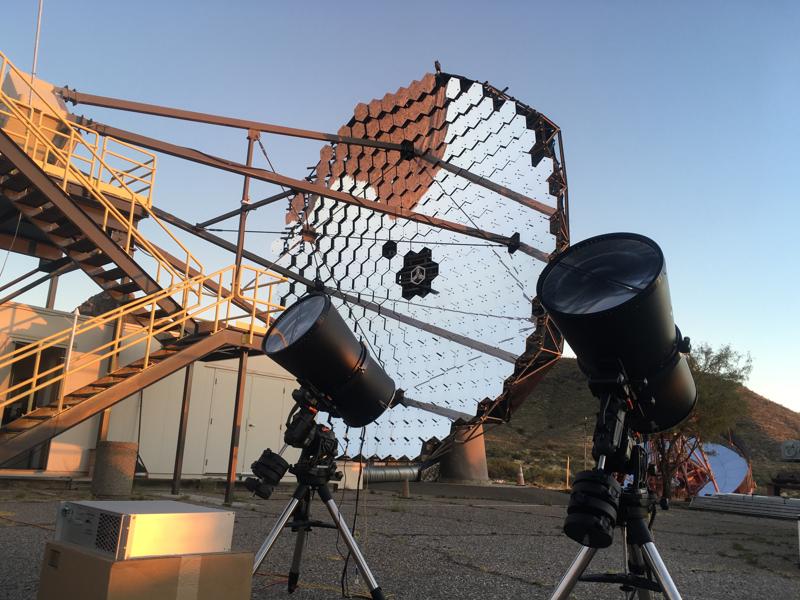}
    \caption{Two PANOSETI telescopes positioned in front of the T4 VERITAS telescope. 
    \label{fig:panoseti}}
\end{figure}

\begin{figure}
    \centering
    \includegraphics[width=0.8\textwidth]{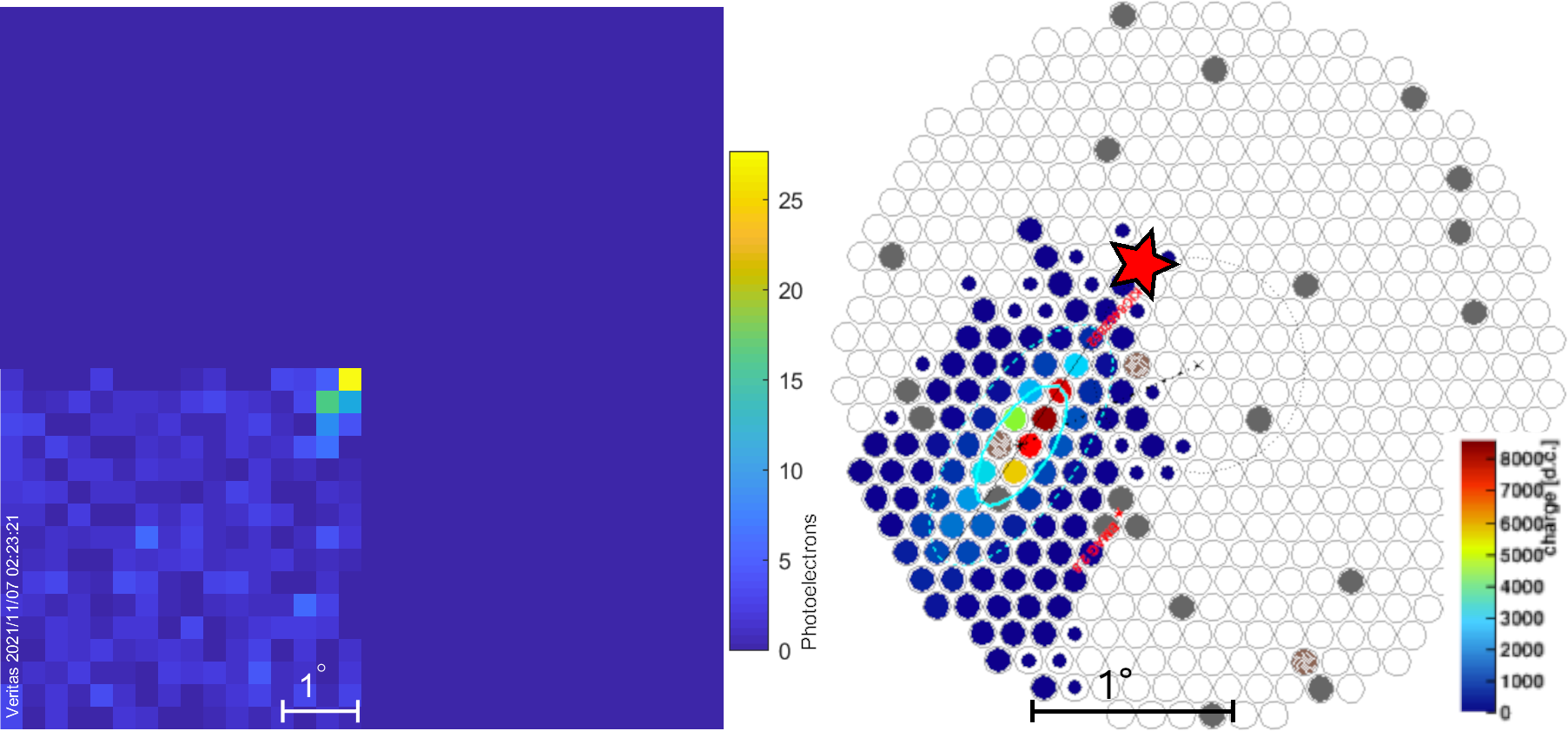}
    \caption{PANOSETI (left) and VERITAS (right) images of a 14.8 TeV shower originating from the Crab Nebula. The Crab Nebula is positioned in the center of the PANOSETI camera, but offset from the center of VERITAS camera at the location of the red star symbol. (A firmware upgrade now triggers all four quadrants when an event is detected.)
    \label{fig:gammaImg}}
\end{figure}

In November 2021, two PANOSETI telescopes were brought to FLWO to observe the Crab Nebula simultaneously with VERITAS, shown in Figure \ref{fig:panoseti} \cite{2022SPIE12184E..8BM}. VERITAS triggers at a rate of 400 Hz, and in 6 hours PANOSETI detected over 10,000 of the showers seen by VERITAS. More than 3,000 of these were seen by both PANOSETI telescopes. At the time of observations, a readout of the PANOSETI camera was triggered only in the quadrant of the camera where any pixel passed the 11.5 photo-electron trigger threshold (updated firmware triggers all four quadrants). Coincident events were identified from the timing information between all telescopes, and a moment analysis was performed on the PANOSETI images to calculate the Hillas parameters for direct comparison of shower geometry with VERITAS \cite{1985ICRC....3..445H, 1997JPhG...23.1013F}.  Measurements of these parameters in the PANOSETI image give a width of 0.13$^\circ$, while VERITAS measures the width to be 0.138$^\circ$. VERITAS could determine that three of the showers seen by both PANOSETI telescopes were very likely from gamma-rays that originated from the Crab Nebula, all with energies above 10 TeV \cite{2022SPIE12184E..8BM}. See Figure \ref{fig:gammaImg} for a comparison of a 14.8 TeV gamma-ray initiated shower between PANOSETI and VERITAS.


\section{Simulation}
\begin{figure}
    \centering
    \includegraphics[width=1.0\textwidth]{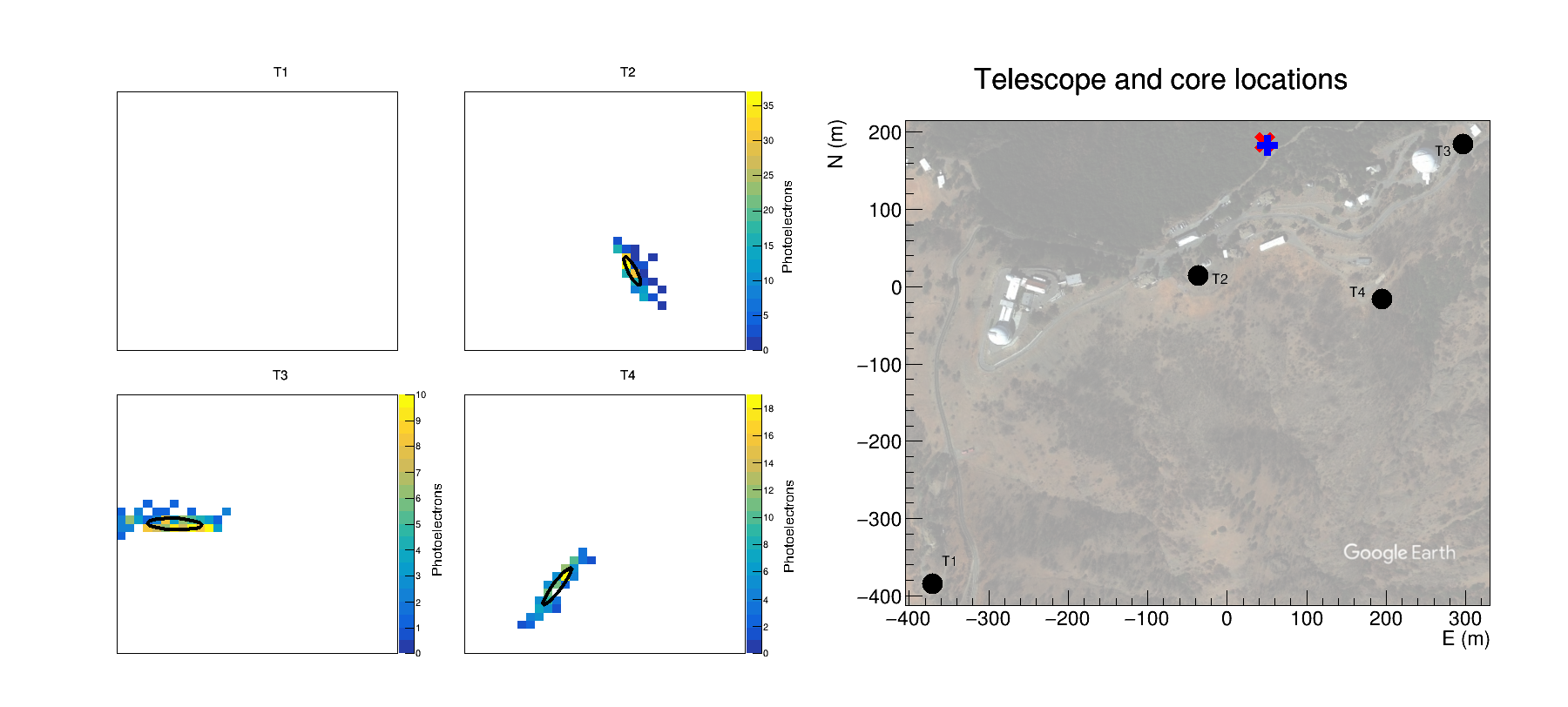}
    \caption{Simulated images of a 100 TeV gamma-ray shower as seen by four PANOSETI telescopes at Lick Observatory (left). The telescopes face North and point to zenith, at the simulated direction of the incident gamma-ray. The approximate locations of the telescopes, with an arbitrary choice of origin, are also plotted (right). The location of the core was reconstructed (blue $+$) to within 5 meters of the true core position (red $\times$), and the direction was reconstructed to 0.03 degrees away from zenith.
    \label{fig:reconstruct}}
\end{figure}

To design an array of PANOSETI telescopes that is optimized for observing gamma-ray showers, we have started developing simulation tools. Air showers are simulated using CORSIKA \cite{1998cmcc.book.....H} version 7.7410, compiled with options typically used for Cherenkov telescopes\footnote{ATMEXT, CERENKOV, IACT, SLANT, and VOLUMEDET} and a combination of hadronic interaction models\footnote{QGSJET-II-04 \cite{Ostapchenko:2010vb} and UrQMD 1.3.1 \cite{Bass:1998ca, Bleicher:1999xi}}. One telescope is simulated at each of the four locations at Lick Observatory mentioned earlier (Figure \ref{fig:reconstruct}, right), with each telescope placed at the same altitude (1239 m above sea level) for simplicity. The atmospheric profile is approximated by measurements taken at VERITAS during winter. The Cherenkov photons are simulated in bunch sizes of 5 to reduce simulation time without compromising image quality \cite{2008APh....30..149B}.

\begin{figure}
    \centering
    \includegraphics[width=0.95\textwidth]{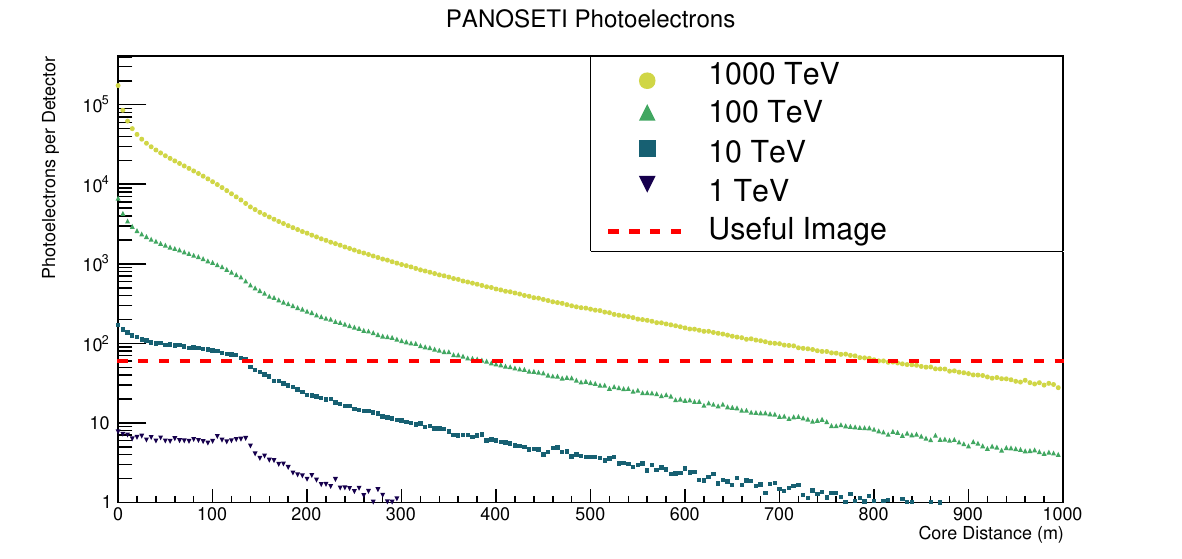}
    \caption{Lateral distribution of Cherenkov light captured by simulated PANOSETI telescopes. Atmospheric extinction and photon detection efficiency is accounted for. Each point is the average of 100 simulated showers. As configured while observing at VERITAS, the PANOSETI telescopes create useful images when more than 60 photo-electrons are detected. Showers with energies below 10 TeV are not bright enough to discern above NSB and cannot be reliably reconstructed. Higher energy showers will appear brighter at the same distance, or can be seen from farther away.
    \label{fig:lateral}}
\end{figure}

The Cherenkov photon output is then read by a modified version of corsikaIOreader \cite{maier_g_2020_4182562}, which is software used to prepare the CORSIKA output for optics simulation and for filling histograms using ROOT \cite{Brun:1997pa}\footnote{See also "ROOT" [software], Release v6.28/04, 30/05/2023, \url{https://root.cern/releases/release-62804/}}. Atmospheric extinction of Cherenkov photons is calculated at this stage rather than in the CORSIKA simulation. The modifications to corsikaIOreader then stochastically attenuate photons to simulate the transmission efficiency of the Fresnel lens and photon detection efficiency (PDE) of the SiPM. The probability of attenuation is calculated using the wavelength dependence of both transmission through PMMA (acrylic) and the SiPM's specified PDE. Figure \ref{fig:lateral} plots the expected number of photo-electrons detected by a PANOSETI telescope as a function of distance to the shower core. 

\begin{figure}
    \centering
    \includegraphics[width=\textwidth]{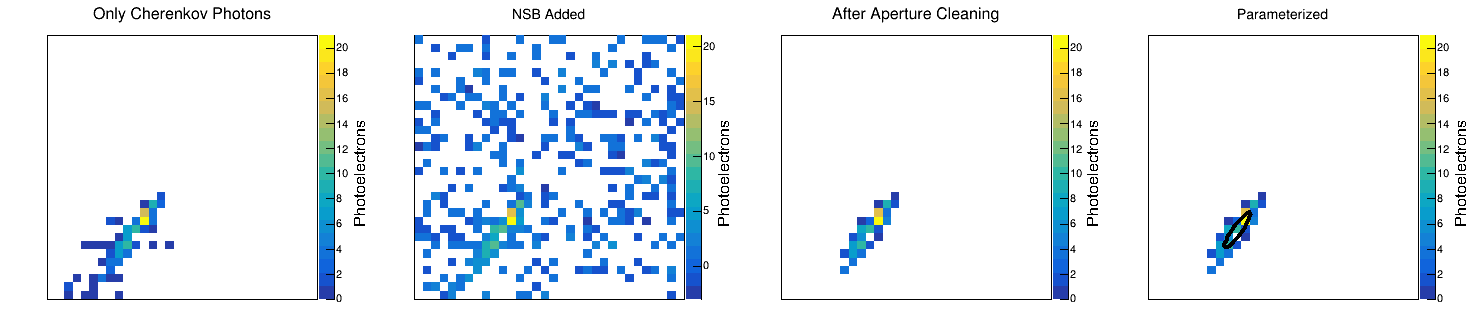}
    \caption{Cherenkov photons from a simulation of a 100 TeV gamma-ray shower in the field of view of a PANOSETI telescope. Each plot shows sequentially (from left to right) the process of adding NSB, applying the aperture cleaning, and parameterizing the cleaned image.
    \label{fig:cleaning}}
\end{figure}

The photon arrival directions are scattered according to the measured optical point spread of the lens \cite{2018SPIE10702E..5LM}. The arrival directions are then used to create images of the shower rather than computing a full ray-trace through a model of the telescope. Night sky background (NSB) is simulated by scaling down the NSB rate measured by VERITAS. Finally, images are cleaned using an aperture cleaning method \cite{2016APh....72...11W}. See Figure \ref{fig:cleaning} for an example of this process. 

To reconstruct the shower direction and core location, Hillas parameters are derived from the cleaned images and used to identify the shower axes. The intersection of shower axes from each unique image pair are averaged together and weighted by the angle between them, based on the method used by HEGRA \cite{1999APh....12..135H}. The Hillas parameterization and shower reconstruction are implemented with routines developed for the VERITAS custom analysis package, Eventdisplay \cite{2017ICRC...35..747M}. Figure \ref{fig:reconstruct} shows an example of reconstruction.

\section{Discussion}
The observations with VERITAS demonstrate that PANOSETI can image Cherenkov showers above 10 TeV. The real image shown in Figure \ref{fig:gammaImg} was reconstructed by VERITAS to land 125 meters away from the telescope, and has a total integrated intensity of 83 photo-electrons in the PANOSETI camera. The intensity predicted by Figure \ref{fig:lateral} of a shower with the same energy and impact parameter is roughly in agreement with what was measured. Observations have also highlighted useful features which are currently under development, such as full camera readout and different triggering modes. Requiring two pixels to pass the trigger threshold instead of one will dramatically reduce the chance of triggering on a fluctuation due to NSB or instrumental artifact, allowing us to lower the required pixel intensity and therefore decrease the energy threshold. Simulations also show that UHE Cherenkov showers can be imaged by PANOSETI without truncation or camera saturation (Figure \ref{fig:reconstruct}), which allows for accurate reconstruction of the arrival direction and energy of the primary photon.

The primary challenge with studying UHE sources is achieving a sufficient effective area. For an IACT, this is determined by the size of the Cherenkov light pool. A single PANOSETI telescope can already detect UHE showers from hundreds of meters away. The four locations at Lick Observatory (Figure \ref{fig:reconstruct}, right) were chosen from sites that have preexisting infrastructure to support deployment. The spacing between the telescopes was chosen to optimize optical observations of SETI targets, and to maximize the amount of light Figure \ref{fig:lateral} predicts would be collected from Cherenkov showers. This pathfinder array is only the first test, and a small array of 24 telescopes spaced a few hundred meters apart can easily fill a footprint that covers a square kilometer. Such an array would provide greater effective area and angular resolution than current experiments for a fraction of the cost. PANOSETI telescopes are much smaller and far more affordable than traditional IACTs. Two were successfully deployed at VERITAS with minimal effort in under a day. An array of PANOSETI telescopes would provide a unique opportunity to resolve the UHE sources, and especially those that cannot be seen at lower energies.

\section{Acknowledgments}
We thank the Lick Observatory staff and engineers for their help in the installation of the PANOSETI telescopes at the Astrograph dome and Barnard sites. We also thank the VERITAS Collaboration and the technical support staff at the Fred Lawrence Whipple Observatory for their cooperation in obtaining joint observations and for the use of their data. The PANOSETI research and instrumentation program is made possible by the enthusiastic support and interest by Franklin Antonio and the Bloomfield Family Foundation. Harvard SETI was supported by The Planetary Society and The Bosack/Kruger Charitable Foundation. UC Berkeley’s SETI efforts involved with PANOSETI are supported by NSF grant 1407804, the Breakthrough Prize Foundation, and the Marilyn and Watson Alberts SETI Chair fund.

\bibliographystyle{JHEP}
\bibliography{bib}

\providecommand{\href}[2]{#2}\begingroup\raggedright\begin{thebibliography}{10}

\bibitem{cao_ultrahigh-energy_2021}
Z.~Cao, F.A.~Aharonian, Q.~An, {Axikegu}, L.X.~Bai, Y.X.~Bai et~al.,
  \emph{Ultrahigh-energy photons up to 1.4 petaelectronvolts from 12
  $\gamma$-ray {Galactic} sources},
  \href{https://doi.org/10.1038/s41586-021-03498-z}{\emph{Nature} {\bfseries
  594} (2021) 33}.

\bibitem{2023arXiv230517030C}
Z.~{Cao}, F.~{Aharonian}, Q.~{An}, {Axikegu}, Y.X.~{Bai}, Y.W.~{Bao} et~al.,
  \emph{{The First LHAASO Catalog of Gamma-Ray Sources}},
  \href{https://doi.org/10.48550/arXiv.2305.17030}{\emph{arXiv e-prints} (2023)
  arXiv:2305.17030} [\href{https://arxiv.org/abs/2305.17030}{{\ttfamily
  2305.17030}}].

\bibitem{Albert_2020}
A.~Albert, R.~Alfaro, C.~Alvarez, J.R.A.~Camacho, J.C.~Arteaga-Velázquez,
  K.P.~Arunbabu et~al., \emph{{3HWC: The Third HAWC Catalog of Very-high-energy
  Gamma-Ray Sources}},
  \href{https://doi.org/10.3847/1538-4357/abc2d8}{\emph{The Astrophysical
  Journal} {\bfseries 905} (2020) 76}.

\bibitem{2018SPIE10702E..5IW}
S.A.~{Wright}, P.~{Horowitz}, J.~{Maire}, D.~{Werthimer}, F.~{Antonio},
  M.~{Aronson} et~al., \emph{{Panoramic optical and near-infrared SETI
  instrument: overall specifications and science program}},  in
  \emph{Ground-based and Airborne Instrumentation for Astronomy VII},
  C.J.~{Evans}, L.~{Simard} and H.~{Takami}, eds., vol.~10702 of \emph{Society
  of Photo-Optical Instrumentation Engineers (SPIE) Conference Series},
  p.~107025I, July, 2018, \href{https://doi.org/10.1117/12.2314268}{DOI}
  [\href{https://arxiv.org/abs/1808.05772}{{\ttfamily 1808.05772}}].

\bibitem{2018SPIE10702E..5LM}
J.~{Maire}, S.A.~{Wright}, M.~{Cosens}, F.P.~{Antonio}, M.L.~{Aronson},
  S.A.~{Chaim-Weismann} et~al., \emph{{Panoramic optical and near-infrared SETI
  instrument: optical and structural design concepts}},  in \emph{Ground-based
  and Airborne Instrumentation for Astronomy VII}, C.J.~{Evans}, L.~{Simard}
  and H.~{Takami}, eds., vol.~10702 of \emph{Society of Photo-Optical
  Instrumentation Engineers (SPIE) Conference Series}, p.~107025L, July, 2018,
  \href{https://doi.org/10.1117/12.2314444}{DOI}
  [\href{https://arxiv.org/abs/1808.05773}{{\ttfamily 1808.05773}}].

\bibitem{2018SPIE10702E..5HC}
M.~{Cosens}, J.~{Maire}, S.A.~{Wright}, F.~{Antonio}, M.~{Aronson},
  S.A.~{Chaim-Weismann} et~al., \emph{{Panoramic optical and near-infrared SETI
  instrument: prototype design and testing}},  in \emph{Ground-based and
  Airborne Instrumentation for Astronomy VII}, C.J.~{Evans}, L.~{Simard} and
  H.~{Takami}, eds., vol.~10702 of \emph{Society of Photo-Optical
  Instrumentation Engineers (SPIE) Conference Series}, p.~107025H, July, 2018,
  \href{https://doi.org/10.1117/12.2314252}{DOI}
  [\href{https://arxiv.org/abs/1808.05774}{{\ttfamily 1808.05774}}].

\bibitem{2015ICRC...34..771P}
N.~{Park} and {VERITAS Collaboration}, \emph{{Performance of the VERITAS
  experiment}},  in \emph{34th International Cosmic Ray Conference (ICRC2015)},
  vol.~34 of \emph{International Cosmic Ray Conference}, p.~771, July, 2015,
  \href{https://doi.org/10.22323/1.236.0771}{DOI}
  [\href{https://arxiv.org/abs/1508.07070}{{\ttfamily 1508.07070}}].

\bibitem{2022SPIE12184E..8BM}
J.~{Maire}, S.A.~{Wright}, J.~{Holder}, D.~{Anderson}, W.~{Benbow}, A.~{Brown}
  et~al., \emph{{Panoramic SETI: program update and high-energy astrophysics
  applications}},  in \emph{Ground-based and Airborne Instrumentation for
  Astronomy IX}, C.J.~{Evans}, J.J.~{Bryant} and K.~{Motohara}, eds.,
  vol.~12184 of \emph{Society of Photo-Optical Instrumentation Engineers (SPIE)
  Conference Series}, p.~121848B, Aug., 2022,
  \href{https://doi.org/10.1117/12.2630772}{DOI}
  [\href{https://arxiv.org/abs/2210.01356}{{\ttfamily 2210.01356}}].

\bibitem{1985ICRC....3..445H}
A.M.~{Hillas}, \emph{{Cerenkov Light Images of EAS Produced by Primary Gamma
  Rays and by Nuclei}},  in \emph{19th International Cosmic Ray Conference
  (ICRC19), Volume 3}, vol.~3 of \emph{International Cosmic Ray Conference},
  p.~445, Aug., 1985.

\bibitem{1997JPhG...23.1013F}
D.J.~{Fegan}, \emph{{TOPICAL REVIEW: $\gamma$/hadron separation at TeV
  energies}}, \href{https://doi.org/10.1088/0954-3899/23/9/004}{\emph{Journal
  of Physics G Nuclear Physics} {\bfseries 23} (1997) 1013}.

\bibitem{1998cmcc.book.....H}
D.~{Heck}, J.~{Knapp}, J.N.~{Capdevielle}, G.~{Schatz} and T.~{Thouw},
  \emph{{CORSIKA: a Monte Carlo code to simulate extensive air showers.}}
  (1998).

\bibitem{Ostapchenko:2010vb}
S.~Ostapchenko, \emph{{Monte Carlo treatment of hadronic interactions in
  enhanced Pomeron scheme: I. QGSJET-II model}},
  \href{https://doi.org/10.1103/PhysRevD.83.014018}{\emph{Phys. Rev. D}
  {\bfseries 83} (2011) 014018}
  [\href{https://arxiv.org/abs/1010.1869}{{\ttfamily 1010.1869}}].

\bibitem{Bass:1998ca}
S.A.~Bass et~al., \emph{{Microscopic models for ultrarelativistic heavy ion
  collisions}},
  \href{https://doi.org/10.1016/S0146-6410(98)00058-1}{\emph{Prog. Part. Nucl.
  Phys.} {\bfseries 41} (1998) 255}
  [\href{https://arxiv.org/abs/nucl-th/9803035}{{\ttfamily nucl-th/9803035}}].

\bibitem{Bleicher:1999xi}
M.~Bleicher et~al., \emph{{Relativistic hadron hadron collisions in the
  ultrarelativistic quantum molecular dynamics model}},
  \href{https://doi.org/10.1088/0954-3899/25/9/308}{\emph{J. Phys. G}
  {\bfseries 25} (1999) 1859}
  [\href{https://arxiv.org/abs/hep-ph/9909407}{{\ttfamily hep-ph/9909407}}].

\bibitem{2008APh....30..149B}
K.~{Bernl{\"o}hr}, \emph{{Simulation of imaging atmospheric Cherenkov
  telescopes with CORSIKA and sim\_ telarray}},
  \href{https://doi.org/10.1016/j.astropartphys.2008.07.009}{\emph{Astroparticle
  Physics} {\bfseries 30} (2008) 149}
  [\href{https://arxiv.org/abs/0808.2253}{{\ttfamily 0808.2253}}].

\bibitem{maier_g_2020_4182562}
G.~Maier, \emph{{corsikaIOreader: a tool to read CORSIKA eventio files}},
  Nov., 2020.
\newblock 10.5281/zenodo.4182562.

\bibitem{Brun:1997pa}
R.~Brun and F.~Rademakers, \emph{{ROOT: An object oriented data analysis
  framework}}, \href{https://doi.org/10.1016/S0168-9002(97)00048-X}{\emph{Nucl.
  Instrum. Meth. A} {\bfseries 389} (1997) 81}.

\bibitem{2016APh....72...11W}
M.~{Wood}, T.~{Jogler}, J.~{Dumm} and S.~{Funk}, \emph{{Monte Carlo studies of
  medium-size telescope designs for the Cherenkov Telescope Array}},
  \href{https://doi.org/10.1016/j.astropartphys.2015.04.008}{\emph{Astroparticle
  Physics} {\bfseries 72} (2016) 11}
  [\href{https://arxiv.org/abs/1506.07476}{{\ttfamily 1506.07476}}].

\bibitem{1999APh....12..135H}
W.~{Hofmann}, I.~{Jung}, A.~{Konopelko}, H.~{Krawczynski}, H.~{Lampeitl} and
  G.~{P{\"u}hlhofer}, \emph{{Comparison of techniques to reconstruct VHE
  gamma-ray showers from multiple stereoscopic Cherenkov images}},
  \href{https://doi.org/10.1016/S0927-6505(99)00084-5}{\emph{Astroparticle
  Physics} {\bfseries 12} (1999) 135}
  [\href{https://arxiv.org/abs/astro-ph/9904234}{{\ttfamily
  astro-ph/9904234}}].

\bibitem{2017ICRC...35..747M}
G.~{Maier} and J.~{Holder}, \emph{{Eventdisplay: An Analysis and Reconstruction
  Package for Ground-based Gamma-ray Astronomy}},  in \emph{35th International
  Cosmic Ray Conference (ICRC2017)}, vol.~301 of \emph{International Cosmic Ray
  Conference}, p.~747, July, 2017,
  \href{https://doi.org/10.22323/1.301.0747}{DOI}
  [\href{https://arxiv.org/abs/1708.04048}{{\ttfamily 1708.04048}}].

\end{thebibliography}\endgroup



%
%
%

\end{document}